\documentclass{article}
\usepackage{TOKsty,amssymb,amsmath,epsfig,amsfonts,textcomp,subfigure,color,epstopdf}
\usepackage[turkish]{babel}
\usepackage[latin5]{inputenc}
\usepackage[T1]{fontenc}
\usepackage{dsfont}
\usepackage{booktabs}
\usepackage{cite}
\usepackage{boxedminipage}
\usepackage{subfigure}
\usepackage{graphicx}
\usepackage{algorithmic}
\usepackage{algorithm}
\usepackage{multirow,bigdelim}
\usepackage{url}
\usepackage{gensymb}
\usepackage{dblfloatfix}
\usepackage{tabularx}
\usepackage{makecell}
\usepackage{textcomp}
\ninept

\title{Kısa Dönem Uzam-Zamansal Trafik Tahmini  \\
	Short-Term Spatio-Temporal Traffic Forecasting }

\usepackage{acronym}

\makeatletter
\def\name#1{\gdef\@name{#1\\}}
\makeatother \name{{\em Akın Taşcıkaraoğlu$^1$, Fatma Yıldız Taşcıkaraoğlu$^2$, İbrahim Beklan Küçükdemiral$^2$}}
\address{$^1$Elektrik Mühendisliği Bölümü   \\
	Yıldız Teknik Üniversitesi, İstanbul \\
	{\small \tt atasci@yildiz.edu.tr}\\ \\
	$^2$Kontrol ve Otomasyon Mühendisliği Bölümü \\
	Yıldız Teknik Üniversitesi, İstanbul \\
	{\small \tt \{fayildiz,beklan\}@yildiz.edu.tr}\\ }
\def\real{ \mathbb{R} }

\newcommand{\vc}[1]{\boldsymbol{#1}}

\acrodef{CoM}{Concentration of Measure}
\acrodef{i.i.d.}{independent and identically distributed}
\acrodef{LTI}{Linear Time-Invariant}
\acrodef{LTV}{Linear Time-Variant}
\acrodef{LPV}{Linear Parameter-Varying}
\acrodef{RIP}{Restricted Isometry Property}
\acrodef{SVD}{Singular Value Decomposition}
\acrodef{CS}{Compressive Sensing}
\acrodef{DSP}{Digital Signal Processing}
\acrodef{CSI}{Compressive System Identification}
\acrodef{CTI}{Compressive Topology Identification}
\acrodef{CBD}{Compressive Binary Detection}
\acrodef{OMP}{Orthogonal Matching Pursuit}
\acrodef{MP}{Matching Pursuit}
\acrodef{ERC}{Exact Recovery Condition}
\acrodef{BOMP}{Block Orthogonal Matching Pursuit}
\acrodef{COMP}{Clustered Orthogonal Matching Pursuit}
\acrodef{CoSaMP}{Compressive Sampling Matching Pursuit}
\acrodef{KKT}{Karush-Kuhn-Tucker}

\acrodef{FIR}{Finite Impulse Response}
\acrodef{DFT}{Discrete Fourier Transform}
\acrodef{DCT}{Discrete Cosine Transform}
\acrodef{JL}{Johnson-Lindenstrauss}
\acrodef{ROC}{Receiver Operating Curve}
\acrodef{NP}{Neyman-Pearson}
\acrodef{ARX}{Auto Regressive with eXternal input} 
\acrodef{MISO}{Multi-Input Single-Output}
\acrodef{SISO}{Single-Input Single-Output}
\acrodef{MIMO}{Multi-Input Multi-Output}
\acrodef{BP}{Basis Pursuit}
\acrodef{LASSO}{Least Absolute Shrinkage and Selection Operator}
\acrodef{GLASSO}{Group LASSO}
\acrodef{NNG}{Non-Negative Garrote}
\acrodef{LARS}{Least Angle Regression}
\acrodef{I/O}{Input/Output}
\acrodef{CST-WSF}{Compressive Spatio-Temporal Wind Speed Forecasting}
\acrodef{CST-SIF}{Compressive Spatio-Temporal Solar Irradiation Forecasting}
\acrodef{CSTF}{Compressive Spatio-Temporal Forecasting}
\acrodef{AR}{Autoregressive}
\acrodef{M-AR}{Multivariate Autoregressive}
\acrodef{NM-AR}{Nonuniform Multivariate Autoregressive}
\acrodef{RPS}{Renewable Portfolio Standard}
\acrodef{NWP}{Numerical Weather Prediction}
\acrodef{TDD}{Trigonometric Direction Diurnal}
\acrodef{RSTD}{Regime Switching Space-Time Diurnal}
\acrodef{TDDGW}{TDD with Geostrophic Wind Information}
\acrodef{ACK}{Nantucket Memorial Airport}
\acrodef{WT}{Wavelet Transform}
\acrodef{ANN}{Artificial Neural Networks}
\acrodef{MAE}{Mean Absolute Error}
\acrodef{MBE}{Mean Bias Error}
\acrodef{RMSE}{Root Mean Squared Error}
\acrodef{NRMSE}{Normalized Root Mean Squared Error}
\acrodef{TSOs}{Transmission System Operators}
\acrodef{IPPs}{Independent Power Producers}
\acrodef{ST-ANN}{Spatio-Temporal ANN}
\acrodef{LS}{Least Squares}
\acrodef{METAR}{Meteorological Terminal Aviation Routine}
\acrodef{PCA}{Principal Component Analysis}
\acrodef{NN}{Neural Network}
\acrodef{PV}{Photovoltaic}
\acrodef{STC}{Standard Test Conditions}
\acrodef{GHI}{Global Horizontal Irradiance}
\acrodef{PCA}{Principal Component Analysis}
\acrodef{SVM}{Support Vector Machines}
\acrodef{GA}{Genetic Algorithm}

\begin{document}
	
	\maketitle
	\addabstract{Özetçe}
	
	Şehir içi yollarda yaşanan tıkanıklık, gecikme ve çevre problemlerini en aza indirmek amacıyla gerçekleştirilen çalışmaların önemi özellikle son yıllarda oldukça artmıştır.
	Bu çalışmalar içerisinde, kısa dönem trafik akımı ve ortalama araç hızı tahmini yöntemleri; uygulanabilirliklerinin kolay olması, farklı amaçlar için etkin bir şekilde kulllanılabilmeleri ve maliyetlerinin oldukça düşük olması nedeniyle ön plana çıkmaktadırlar.
	Trafik ağındaki istenen noktalara ait bağ akımlarının ve araç hızlarının gelecekteki olası değerlerinin tahmin edildiği bu yöntemlerin, trafik yönetiminde yaşanabilecek sorunları öngörerek trafik sıkışıklığını azaltmaya yardımcı olduğu çok sayıda çalışmada belirtilmektedir. 
	Bu yayında, bir bağdaki ortalama araç hızı tahminleri için çok sayıda noktadan alınan geçmiş verileri göz önüne alan bir \textit{uzam-zamansal} yaklaşım sunulmaktadır. 
	Belirtilen yaklaşım, her bir aşamada en faydalı verileri belirleyerek yalnızca bu verilerin giriş veri kümesine dahil edilmesini sağlayan bir algoritma içermektedir. 
	Algoritma içerisinde \textit{seyrek} matrislerin elde edilmesi ile tahmin doğruluğunun arttırılabilmesi sağlanırken tahminlerin gerçekleştirilme sürelerinin de literatürde yer alan yöntemlere oranla kayda değer derecede iyileştirilmesi hedeflenmektedir.   
	
	\addabstract{Abstract}
	The studies carried out with the objective of minimizing the effects of congestion, delay and environment problems on the transportation network have gained increasing importance in the last years. 
	Among these studies, short-term traffic flow and average vehicle speed forecasting methods have come into prominence due to their easy implementations, efficient usage on different areas and cost-effectiveness.     
	A large number of studies have reported that these methods, in which the expected future values of link flows and average speeds are forecasted in desired points, can reduce the traffic congestion by anticipating the problems in traffic management. 
	In this paper, a \textit{spatio-temporal} approach accounted for historical traffic characteristics data collected from a large number of points is presented for average speed forecasts in a given link. 
	The proposed approach includes an algorithm that enables to take into account the most informative data in an input set by determining them for each stage.
	It is aimed to increase the forecasting accuracy by using \textit{sparse} matrices in the algorithm while decreasing the calculation times significantly compared to the similar methods presented in the literature.
	
	\section{Giriş}
	Gelişmiş ülkelerdeki büyük kentlerde yaşanan nüfus ve araç sahipliğindeki hızlı artışın ve şehirleşmenin ulaştırma ağı üzerindeki etkilerine bağlı olarak yaşanan tıkanıklık, gecikme, güvenlik ve çevre problemlerini en aza indirmek amacıyla gerçekleştirilen ve bilişim teknolojileriyle desteklenen faaliyetler, Akıllı Ulaşım Sistemleri olarak adlandırılmaktadır. 
	Şehirlerde yaşanan trafik tıkanıklığının insanların yaşam kalitesi üzerindeki doğrudan etkileri nedeniyle akıllı ulaşım sistemlerinin, gelecekteki akıllı şehir planlamalarında çok önemli bir bileşen olarak yer alması beklenmektedir. 
	Bu nedenle, trafiğin durumundaki beklenen gelişmeler, trafiğin güvenli ve ekonomik bir şekilde yönetilmesi açısından oldukça büyük bir öneme sahiptir.
	Belirli trafik karakteristiklerinin önceden bilinmesi, trafik yönetim sistemlerine veri desteği sağlamalarının yanı sıra kullanıcılara da trafik şartları hakkında önemli bilgiler verebilir.
	Ancak sürekli değişen yapısı nedeniyle, trafik tıkanıklığı ve tıkanıklığı etkileyen faktörler arasındaki etkileşimleri modellemek için gelişmiş yöntemlere ihtiyaç duyulmaktadır. 
	
	Bu amaçla geliştirilen tahmin yöntemlerinin yapısı öncelikle yolun türüne (otoyol, şehir içi arter, vb.) bağlıdır. 
	Yol türlerinin her birinin kendine özgü bir karakteristiği vardır. 
	Örneğin, ışıklı kavşaklar ve yaya geçişleri gibi çeşitli kısıtlar, otoyollara oranla şehir içi arterlerdeki tahmin işlemini daha karmaşık bir hale getirmektedir~\cite{tascikaraoglu2015pointq}. 
	Şehir merkezlerine yakın olan ve çok sayıda katılıma sahip olan otoyollardaki trafik karakteristiklerinin tahminleri de benzer şekilde şehirlerarası otoyollara kıyasla daha zor bir işlem olarak nitelendirilebilir.       
	Model seçimindeki bir diğer önemli faktör ise gerekli tahmin aralığıdır ve bu süre çoğunlukla tahminlerin kullanılacağı uygulama alanı göz önüne alınarak belirlenmektedir. 
	Bir genelleme yapacak olursak, yüksek doğrulukta modellemeye olanak sağlamalarından dolayı, daha küçük tahmin adımlarının gerçek uygulamalar için daha değerli olduğu söylenebilir.  
	
	Akım, yolculuk zamanı, kuyruk uzunluğu, hız ve yoğunluk gibi farklı trafik karakteristikleri arasında trafik akımı ve hız tahminleri, trafik tıkanıklığının azaltılması ve akıllı ulaşım sistemlerinde ulaşım hareketliliğinin arttırılması açısından anahtar bir role sahiptir. 
	Özellikle, birkaç dakikadan birkaç saate kadar olan kısa dönem trafik akımı ve ortalama hız tahminleri, dinamik trafik kontrolünü destekleyerek, gecikmeleri ve trafik tıkanıklığını azaltmada oldukça etkili olmaktadırlar. 
	Belirtilen parametrelerin tahmininde kullanılan klasik yöntemler kararlı trafik şartları için belli bir doğrulukta tahminler sağlamaktadırlar. 
	Ancak, trafiğin en yoğun olduğu saatlerdeki yüksek taşıt sayısı, yetersiz yollar, kazalar, koordinasyonu sağlanamamış trafik ışıkları ve kötü hava şartları gibi çok sayıda etkenin neden olduğu yoğun trafik durumları için bu yöntemler çoğunlukla hatalı tahminler vermektedirler. 
	
	Birbirine fiziksel olarak bağlı yollardan alınan veriler, trafik şartlarındaki bu kısa ve uzun süreli değişimleri dikkate alarak tahmin yöntemlerinin doğruluğunu iyileştirebilirler. 
	Özellikle, trafiğin akış yönüne göre daha başlarda bulunan bağlardan alınan verilerin, sonraki bağların tahmini üzerinde genellikle güçlü bir etkisi olacaktır. 
	Bu amaçla, bir ön araştırma gerçekleştirilerek belirli bir ağdaki trafik akımına ve ortalama hıza en fazla etki eden faktörler belirlenebilir ve yalnızca bu faktörler tahmin algoritmalarında kullanılabilir. 
	Alternatif olarak, tıkanık ve tıkanık olmayan trafik şartları için çoklu rejim (multi-regime) yöntemleri de tercih edilebilir. 
	Ancak, trafik akımı ile trafik şartlarındaki beklenen ve beklenmeyen olaylar arasındaki karmaşık ilişkinin tam olarak belirlenebilmesi genellikle mümkün olmamaktadır. 
	Üstelik bu ilişki zamanla sürekli olarak değişmektedir ve böylece bir model elde etmek daha da zor hale gelmektedir. 
	Bu sorunların üstesinden gelmek amacıyla, ağ üzerindeki çeşitli noktalardaki belli bir sayıda aday değişken içerisinden her bir adımda en uygun girişleri belirleyen ve bir sonraki tahminde yalnızca bu verileri dikkate alan tahmin yaklaşımları literatürde son yıllarda oldukça ön plana çıkmıştır. 
	\textit{Uzam-zamansal} modeller olarak adlandırılan bu gelişmiş modeller, geniş bir ağ üzerindeki çeşitli bilgileri dikkate almaları nedeniyle, trafikteki değişen şartlara kolaylıkla uyum sağlayabilirler. 
	Bu nedenle, tekrarlanmayan geçici şartlar dahil farklı trafik şartları için sürdürülebilir bir tahmin kalitesi sağlayabilirler.
	
	Literatürde özellikle son yıllarda, trafik ağlarındaki verilerin toplanması için kullanılmaya başlanan gelişmiş teknolojiler ile birlikte geniş bir aralıkta uzamsal ve zamansal trafik verilerinin elde edilebilmesi, trafik akımı ve ortalama araç hızı tahmini alanında bu verileri kullanan çok sayıda yöntemin geliştirilmesine olanak sağlamıştır. 
	Stathopoulos vd.~\cite{Stathopoulos2003multivariate} bir şehir içi ışıklı kavşak için çok değişkenli zaman serileri modellerini kullanarak kısa dönemli akım tahminleri gerçekleştirmişlerdir. 
	Benzer yapıdaki bir diğer çok değişkenli uzam-zamansal model ise Dong vd.~\cite{dong2014flow} tarafından geliştirilerek, tıkanık ve tıkanık olmayan trafik durumlarında hız ve akım tahminleri için modelin etkinliği ayrı ayrı test edilmiştir. 
	Sun vd.~\cite{sun2006bayesian} ve Tselentis vd.~\cite{tselentis2014improving} uzamsal ve zamansal trafik akımı tahminlerinde Bayesian ağları yöntemini kullanmışlardır. 
	Bulanık mantık tabanlı bir şehir içi trafik tahmini yöntemi ise Dimitriou vd.~\cite{dimitriou2008adaptive} tarafından önerilmiştir. 
	Bu yayında önerilen tahmin yöntemine benzer olarak, Sun ve Zhang~\cite{sun2007selective} bir ağdaki bağ akımları arasındaki korelasyonları dikkate alarak belirli bir ağ için yalnızca yüksek korelasyona sahip ağlardan gelen verileri giriş verisi kümesine dahil etmişlerdir. 
	Sonuç olarak özellikle birbirine yakın olan ağlardan alınan verilerin tahmin doğruluğu üzerinde olumlu etkisi olduğunu belirtmişlerdir. 
	Ancak yalnızca belirli ağlardan alınan verilerin tahmin algoritmalarında kullanılması, trafikteki aniden meydana gelen beklenmeyen durumlarda çoğunlukla başarısız tahminler üretmektedirler. 
	Bu amaçla bu yayında önerilen yöntem içerisinde, bağlar arasındaki korelasyonlar sürekli olarak hesaplanarak yalnızca mevcut durum için en yüksek korelasyona sahip bağlardan gelen veriler dikkate alınmaktadır. 
	Trafik karakteristiklerine ait verilerin model içerisine dahil edilmesi işlemi her bir tahmin ufkunda tekrar değerlendirilmektedir ve bu sayede yalnızca bazı bağların tahminler üzerindeki etkilerini sürekli olarak kullanan literatürdeki modellerden~\cite{dunne2013weather} daha yüksek tahmin doğruluklarının elde edilmesi hedeflenmektedir. 
	Uzamsal verinin trafik tahminindeki kullanımı ve önemi Vlahogianni vd.~\cite{vlahogianni2014short} tarafından geniş bir çerçevede incelenmiştir.  
	
	Bu çalışmada, uzamsal ve zamansal trafik akım hızı tahmini yöntemlerinin, özellikle ortalama yolculuk zamanlarında ve tüketilen yakıt miktarlarında sağladığı iyileşmeler göz önüne alınarak, yüksek tahmin doğruluğuna sahip bir uzam-zamansal ortalama hız tahmini yöntemi geliştirilmiştir. 
	Tahmin modeli içerisine farklı bağlara ait verilerin dahil edilebilmesi için, matematiksel olarak basit bir çözüm sunmaları nedeniyle çok değişkenli zaman serileri modelleri kullanılmıştır. 
	Önerilen yöntemin iki ana katkısının olacağı düşünülmektedir.
	Trafik akım hızlarının gelecekteki değerlerinin tahmin edilmesi ve bu sayede trafik tıkanıklığı açısından sorun yaşaması muhtemel olan bölgelerde, trafik ışıklarının yeşil sürelerine müdahale edilmesi ve/veya taşıt sahiplerinin trafik tabelaları ve akıllı telefon uygulamaları (Google maps, Yandex navigator, vb.) ile bilgilendirerek alternatif yollara yönlendirilmesi, hedeflenen katkılardan ilkini oluşturmaktadır. 
	Diğer katkı ise özellikle trafikte yaşanacak geçici sorunların çözümünde yöntemin sağlayacağı faydadır. 
	Örneğin, literaturde geleneksel olarak kullanılan modelleme ve parametre kestirimi yöntemleri ile zamanlama planı belirlenen bir yolda meydana gelebilecek bir trafik kazası, yapılan planlamaların etkinliğini önemli ölçüde azaltacaktır. 
	Bu ve trafik akımındaki benzeri geçici sorunların etkileri, yapılacak olan tahminler ile belirli bir süre önce öngörülerek en aza indirilebilir. 
	Modelleme yaklaşımlarının aksine tahmin yöntemleri dinamik olarak güncellenen veri kümelerini kullandıkları için bahsedilen geçici sorunların etkilerinin belirlenmesinde literatürde sıklıkla kullanılmaktadırlar. 
	Mevcut yayın kapsamında geliştirilen tahmin yönteminin, hem modelleme ve parametre kestirimi yaklaşımları ile bir arada kullanılabileceği hem de geçici durumlarda bu iki yönteme alternatif olabileceği söylenebilir. 
	Her iki durumda da tek amaç, trafik kontrol yöntemlerinin etkinliğini en üst seviyeye çıkarmaktır. 
	Bu sayede, mevcut trafik ağları kullanılarak, insanların trafikte geçirdiği süreleri en aza indirebilmek, trafiğin insanlar üzerindeki sosyal ve ekonomik etkilerini azaltabilmek, taşıtların neden olduğu karbon emisyonlarının miktarını düşürebilmek ve trafiğin neden olduğu diğer tüm olumsuz etkileri azaltabilmek daha mümkün hale gelecektir.

\section{Uzam-Zamansal Tahmin Modeli}

\begin{figure}[h!]
	\centering
	\resizebox{.45\textwidth}{!}{\includegraphics{{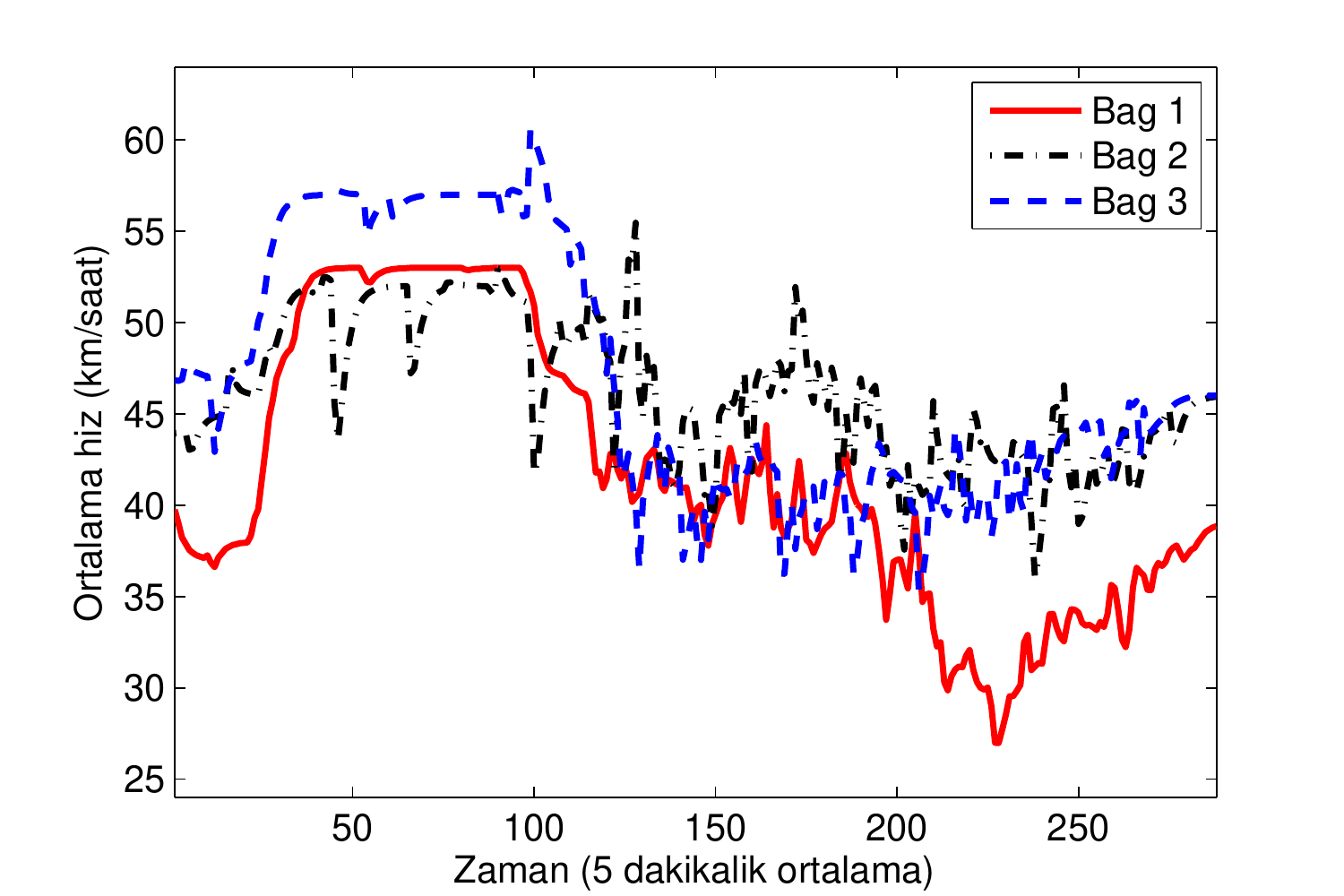}}}
	\caption{Farklı bağlara ait ortalama hız ölçümleri.}
	\label{similarity}
\end{figure}

Bu yayında bir kısa dönem ortalama hız tahmini yöntemi geliştirilmiştir.
Geliştirilen yöntem, önceki bölümde belirtildiği gibi farklı noktalardan alınan veriler arasındaki korelasyonların dikkate alınması prensibine dayanmaktadır.
Şekil~\ref{similarity}'de bu yayında kullanılan veri seti içerisindeki üç farklı bağa ait beş dakika ortalamalı hız verileri gösterilmektedir.
Bir güne ait verilerin yer aldığı bu şekilde birbirine yakın farklı yollardaki ortalama hız değerlerinin benzer karakteristiğe sahip olduğu açıkça görülmektedir. 
Üç bağa ait değerler bazen aynı anda azalırken veya artarken bazı zamanlarda ise bu değişimler belli bir zaman gecikmesi ile diğer bağlarda gözlemlenmektedir.
Özellikle zaman gecikmelerine sahip bilgilerin önerilen tahmin yönteminin etkinliği üzerinde oldukça önemli bir etkisi bulunmaktadır.

Geliştirilen yöntemin temeli, yüksek doğrulukta tahminler sağlaması ve basit yapısı nedeniyle literatürde kısa süreli tahminler için sıklıkla kullanılan Otoregresif Modele (Autoregressive - AR) dayanmaktadır~\cite{Tascikaraoglu2014review}. 
Bu modeller ayrıca, geliştirilecek olan modelin ana hedefi olan, çok sayıda bağdan alınan ve farklı trafik karakteristiklerine ait olan tüm verilerin model içerisine dahil edilmesi fikrine olanak sağlamaktadır. 
AR modeller, bir sistemin çıkış değişkeninin, sisteme ait geçmiş verilerin ağırlıklandırılmış doğrusal bir kombinasyonu olarak elde edilebileceği varsayımına dayanırlar. 
Bu yaklaşımı, yayındaki geniş bir veri setinin kullanılması hedefine uygun olarak çok değişkenli sistemlere uygulayabiliriz. 
Burada $P$ ($p = 1, 2, \dots, P$) adet çıkış değişkenine (örneğin, $P$  adet bağın ortalama hız değerlerine) ait ölçümlere sahip olduğumuzu varsayalım.
Bu çıkış değişkenleri, $P = V \times S$ olmak üzere $S$ ($s = 1, 2, \dots, S$) adet düğümden alınan $V$ ($v = 1, 2, \dots, V$) adet değişkene ait ölçümler olabilir.  
$t$ ($t = 1,2,\dots,M+n$) örnek zamanında $s$'inci bağda ölçülen  $v$'ninci değişkenin ölçümünü $y^{v^{*},s^{*}}$ olarak tanımlayalım ve bu durumda hedef (tahmin edilmek istenen) çıkış değişkeni $y^{v,s}_t$ olsun. 
Sonuç olarak bir Çok Değişkenli AR model (Multivariate Autoregressive - M-AR), diğer adıyla Vektör Otoregresif model~\eqref{M_AR_model}'deki gibi elde edilebilir:

\begin{equation}
y^{v^{*},s^{*}}_t = \sum_{\substack{v=1\\ s=1}}^{V,S} \sum_{i=1}^n y^{v,s}_{t-i} x^{v,s}_i + e^{v^{*},s^{*}}_t,
\label{M_AR_model}
\end{equation}

\begin{figure*}[h!]
\begin{equation}
\small
\underbrace{
\left[
\begin{array}{c}
y_{n+1}^{v^{\star},s^{\star}}\vspace{0.05in}\\
y_{n+2}^{v^{\star},s^{\star}}\\
\vdots\\
y_{n+M}^{v^{\star},s^{\star}}
\end{array}
\right]}_{\vc{b}\in \real^{M}}=
\underbrace{
\left[
\begin{array}{ccc|cc|ccc}
y_{n}^{1,1} & \hdots & y_{1}^{1,1} & \hdots & \hdots &y_{n}^{V,S} & \hdots & y_{1}^{V,S}\\
y_{n+1}^{1,1}  & \ddots & \vdots & & & y_{n+1}^{V,S} & \ddots & \vdots \\
\vdots & \ddots & \vdots & & & \vdots & \dots & \vdots\\
y_{n+M-1}^{1,1} &  \hdots & y_{M}^{1,1} & \hdots & \hdots & y_{n+M-1}^{V,S}  & \hdots & y_{M}^{V,S} \\
\end{array}
\right]}_{\vc{A} \in \real^{M \times N}}
\underbrace{
\left[
\begin{array}{c}
x_1^{1,1}\\
\vdots\\
x_n^{1,1}\\[1ex]
\hline \\[-1.5ex]
\vdots\\
\vdots\\[1ex]
\hline \\[-1.5ex]
x_1^{V,S}\\
\vdots\\
x_n^{V,S}\\
\end{array}
\right]}_{{\vc{x} \in \real^{N}}}
\begin{array}{l}\\[-12mm] \rdelim\}{4}{6mm}[$\text{Blok}~1$] \\ \\ \\ \\ \\ \\[7mm] \rdelim\}{4}{6mm}[$\text{Blok}~P$] \\ \\
\end{array} \\[-1ex]
\hspace{0.2in}
+
\underbrace{
\left[
\begin{array}{c}
e_{n+1}^{v^{\star},s^{\star}}\vspace{0.05in}\\
e_{n+2}^{v^{\star},s^{\star}}\\
\vdots\\
e_{n+M}^{v^{\star},s^{\star}}
\end{array}
\right]}_{\vc{e}\in \real^{M}}
\label{X_vector}
\end{equation}
\end{figure*}
\begin{figure*}[t!]
	\begin{equation}
	\small
	\underbrace{
		\left[
		\begin{array}{c}
		y_{n_{\text{max}}+1}^{v^{\star},s^{\star}}\vspace{0.05in}\\
		y_{n_{\text{max}}+2}^{v^{\star},s^{\star}}\\
		\vdots\\
		y_{n_{\text{max}}+M}^{v^{\star},s^{\star}}
		\end{array}
		\right]}_{\vc{b}\in \real^M}=
	\underbrace{
		\left[
		\begin{array}{ccccc|c|ccc}
		y_{n_{\text{max}}}^{1,1} & \hdots & \hdots & \hdots & y_{n_{\text{max}}-n_1+1}^{1,1} & \hdots &y_{n_{\text{max}}}^{V,S}& \hdots & y_{n_{\text{max}}-n_P+1}^{V,S}\\
		y_{n_{\text{max}}+1}^{1,1}  & \ddots & \ddots & \ddots & \vdots & &y_{n_{\text{max}}+1}^{V,S} & \ddots & \vdots \\
		\vdots & \ddots & \ddots & \ddots & \vdots & & \vdots & \ddots & \vdots\\
		y_{n_{\text{max}}+M-1}^{1,1} &  \hdots  & \hdots & \hdots & y_{n_{\text{max}}-n_1+M}^{1,1} & \hdots & y_{n_{\text{max}}+M-1}^{V,S} & \hdots &y_{n_{\text{max}}-n_P+M}^{V,S} \\
		\end{array}
		\right]}_{\vc{A} \in \real^{M \times N}}
	\underbrace{
		\left[
		\begin{array}{c}
		x_1^{1,1}\\
		\vdots\\
		\vdots\\
		\vdots\\
		x_{n_1}^{1,1}\\[1ex]
		\hline \\[-1.5ex]
		\vdots\\[1ex]
		\hline \\[-1.5ex]
		x_1^{V,S}\\
		\vdots\\
		x_{n_P}^{V,S}\\
		\end{array}
		\right]}_{{\vc{x} \in \real^N}}
	\begin{array}{l}\\[-13mm] \rdelim\}{7}{6mm}[$\text{Blok}~1$] \\ \\ \\ \\ \\[19mm] \rdelim\}{4}{6mm}[$\text{Blok}~P$] \\ \\
	\end{array} \\[-1ex]
	+
	\underbrace{
		\left[
		\begin{array}{c}
		e_{n_{\text{max}}+1}^{v^{\star},s^{\star}}\vspace{0.05in}\\
		e_{n_{\text{max}}+2}^{v^{\star},s^{\star}}\\
		\vdots\\
		e_{n_{\text{max}}+M}^{v^{\star},s^{\star}}
		\end{array}
		\right]}_{\vc{e}\in \real^M}
	\label{X_vector_nonuniform}
	\end{equation}
\end{figure*}
Burada $x^{v,s}_i$ ($\forall i, v, s$) ilgili regresyon katsayılarını, $n$ model derecesini ve $e^{v^{*},s^{*}}_t$ Gauss gürültüsünü (Gaussien noise) temsil etmektedir. 
Eşitlik~\eqref{M_AR_model}, $N := n P$ olacak şekilde Eşitlik~\eqref{X_vector}'de gösterildiği gibi genişletilebilir. 
Model eğitimi aşamasında amaç, $e$ gürültü bileşenini de dikkate alarak $\vc{b} \in \real^M$ ve $\vc{A}\in \real^{M \times N}$ gözlemlerini en iyi şekilde açıklayan bir $\vc{x}\in\real^N$ katsayı vektörünü belirlemektir. 
Eşitlik~\eqref{X_vector}'den açıkça görülebileceği üzere $x$ katsayıları bir blok yapı içerisinde toplanmışlardır. 
Diğer bir ifadeyle, her bir değişkene ait katsayılar $n$ uzunluğundaki bir blok vektöre karşılık gelmektedir.	

Farklı trafik karakteristiklerine ait olan ve farklı noktalardan ölçülen değerler birbirinden farklı büyüklüklere sahip olduklarından, $b$ vektörü ve $A$ matrisindeki tüm değerler bir normalizasyon işlemi ile 0-1 aralığındaki değerlere dönüştürülmektedir. 
Bu sayede, bu değerlerden herhangi birinin $\vc{x}$ vektörü katsayılarının belirlenmesindeki etkisinin diğer değişkenlere oranla çok daha fazla veya az olması önlenmektedir.

Eşitlik~\eqref{X_vector}'de tüm $n$ derece değerlerinin eşit olduğu varsayılmıştır. 
Başka bir ifadeyle, hedef bağ akımı ve diğer bağlardaki değişkenlerin aynı $n$ derecesindeki AR modelleri ile ilişkilendirildiği kabul edilmiştir. 
Ancak her bağın ve bağlara ait her değişkenin (akım, yolculuk süresi, kuyruklanma uzunluğu, araç hızları, yoğunluk, vb.) tahmin değerlerine olan etkisinin farklı olacağı açıktır. 
Bu nedenle Eşitlik~\eqref{X_vector}'de verilen M-AR modelinin farklı $n$ derecelerine sahip değişkenler için düzenlenmiş hali, $n_i$ ($i=1,2,\dots,P$) $i$'ninci bağa ait derece, $n_{\text{max}} \geq \max_i n_i$ ve $N := \sum_{i=1}^Pn_i$ olmak üzere Eşitlik~\eqref{X_vector_nonuniform}'de verilmiştir. 
Bu modelde her biri farklı blok uzunluklarına sahip olan düzgün dağılmamış blok katsayısı matrisi $x$ elde edilir. 
Bu sayede hedef akım çıkışı ile her bir bağa ait değişkenler arasındaki korelasyonlara göre tüm değişkenler arasında bir ayrım yapılması sağlanmıştır.

Verilen modellerde $x$ vektör katsayıları farklı yollarla elde edilebilir. 
Örneğin en küçük kareler yöntemi kullanılarak her bir tahmindeki hataların karelerinin toplamı en küçük olacak şekilde gerekli katsayılar elde edilebilir. 
Ancak hız tahminleri için geliştirilecek olan model çok sayıda farklı bağın her birinden alınacak olan trafik karakteristiklerine ait verileri içerecektir. 
Dolayısıyla tahminlerin gerçekleştirilebilmesi için harcanacak süre kısa dönemli bir tahmin algoritması için oldukça fazla olacaktır. 
Kısa süreli tahminlerde, tahmin süresinin kayda değer derecede artması, elde edilen yüksek tahmin doğruluklarının önemini azaltmaktadır. 
Bu nedenle, bu yayında geliştirilen modelde kullanılacak yöntemlerle seyrek (sparse) $x$ vektörleri elde edilmeye çalışılmaktadır. 
Burada bir varsayım yapılarak çok sayıda bağdan alınan hız verileri arasında yalnızca bazılarının, tahmin edilmeye çalışılan bağa ait ortalama hız ile arasında güçlü bir korelasyon olduğu göz önüne alınmaktadır.
Fiziksel olarak da anlamlı olacak olan bu belirli değerlerin tahmin üzerindeki olumlu etkileri düşünülerek tahmin aşamasında yalnızca bu değerlerin dikkate alınması sağlanmaktadır.
Başka bir ifade ile tahmin doğruluğundan ödün vermeksizin $x$ vektör elemanlarının mümkün olduğunca büyük bir kısmının sıfır olması sağlanmaktadır. 
Literatürde bu vektörler blok-seyrek (block-sparse) olarak adlandırılmaktadır. 
Burada en önemli soru sınırlı sayıdaki sıfır olmayan $x$ vektörü elemanlarının nasıl belirleneceğidir. 
Literatürde rüzgar hızının tahmin edildiği çalışmalarda (~\cite{sanandaji2015low,tascikaraoglu2016exploiting}), sıkıştırmalı algılama (Compressive Sensing) yöntemi kullanarak hesaplanan $x$ vektörleri sayesinde tahmin doğruluğu açısından elde edilen kayda değer sonuçlar göz önüne alınarak bu yayında blok-seyrek $x$ vektörleri \eqref{find_eq}'te verilen optimizasyon problemi çözülerek hesaplanmıştır.

\begin{equation}
\min_{\vc{x}} \|\vc{b}-\vc{A}\vc{x}\|_2 \ \ \ \ \ \ (\vc{x} \ \text{blok-seyrek olmak üzere}).
\label{find_eq}
\end{equation}
%


\section{Benzetim Çalışmaları}
\label{case}

Önerilen yöntemin etkinliğini ölçmek amacıyla literatürde çok sayıda yayında yer verilen Kaliforniya Otoyol Performans Ölçüm Sistemine (PEMS v14.1) ait bir veri seti kullanılmıştır.
Bu sistem Kaliforniya'da bulunan otoyollar üzerindeki çok sayıdaki noktadan gerçek zamanlı verileri kaydetmektedir.
İnternet üzerinden erişilebilen sayfasında \cite{pems} bu veriler beşer dakikalık çözünürlükte sağlanmaktadır.
Literatürdeki ilgili yayınların çok büyük bir kısmında bu veriler yarım saat veya bir saatlik ortalamalara dönüştürülerek kullanılmaktadırlar.
Bu çalışmaların temel hedefi genellikle daha yüksek zamansal çözünürlüğe sahip veriler kullanarak daha uzun süreli tahminler gerçekleştirmektedir.
Ayrıca düzeltme etkisi (smoothing effect) nedeniyle uzun sürelerde daha kararlı değerler elde edilmektedir ve böylece daha yüksek tahmin doğrulukları sağlanmaktadır.
Bu çalışmada ise beş dakikalık veriler doğrudan kullanılarak 30 dakikaya kadar ortalama hız değerleri tahmin edilmiştir. 
Tahminler için 9 saatlik (5:00-14:00 arası 108 adet beş dakikalık değer) bir eğitim seti kullanılarak aynı gün içerisindeki bu zamanı takip eden 9 saatin (14:00-23:00 arası degerler) tahmini gerçekleştirilmiştir. 
Özellikle kısa süreli trafik bilgisine ihtiyaç duyulan uygulamalarda (navigasyon yazılımları, trafiğin durumunu gösteren dinamik trafik tabelaları, vb.) bu yöntemin çok daha etkin olacağı düşünülmektedir.

\begin{figure}[h!]
	\centering
	\subfigure[AR Model]{
		\resizebox{.45\textwidth}{!}{\includegraphics{{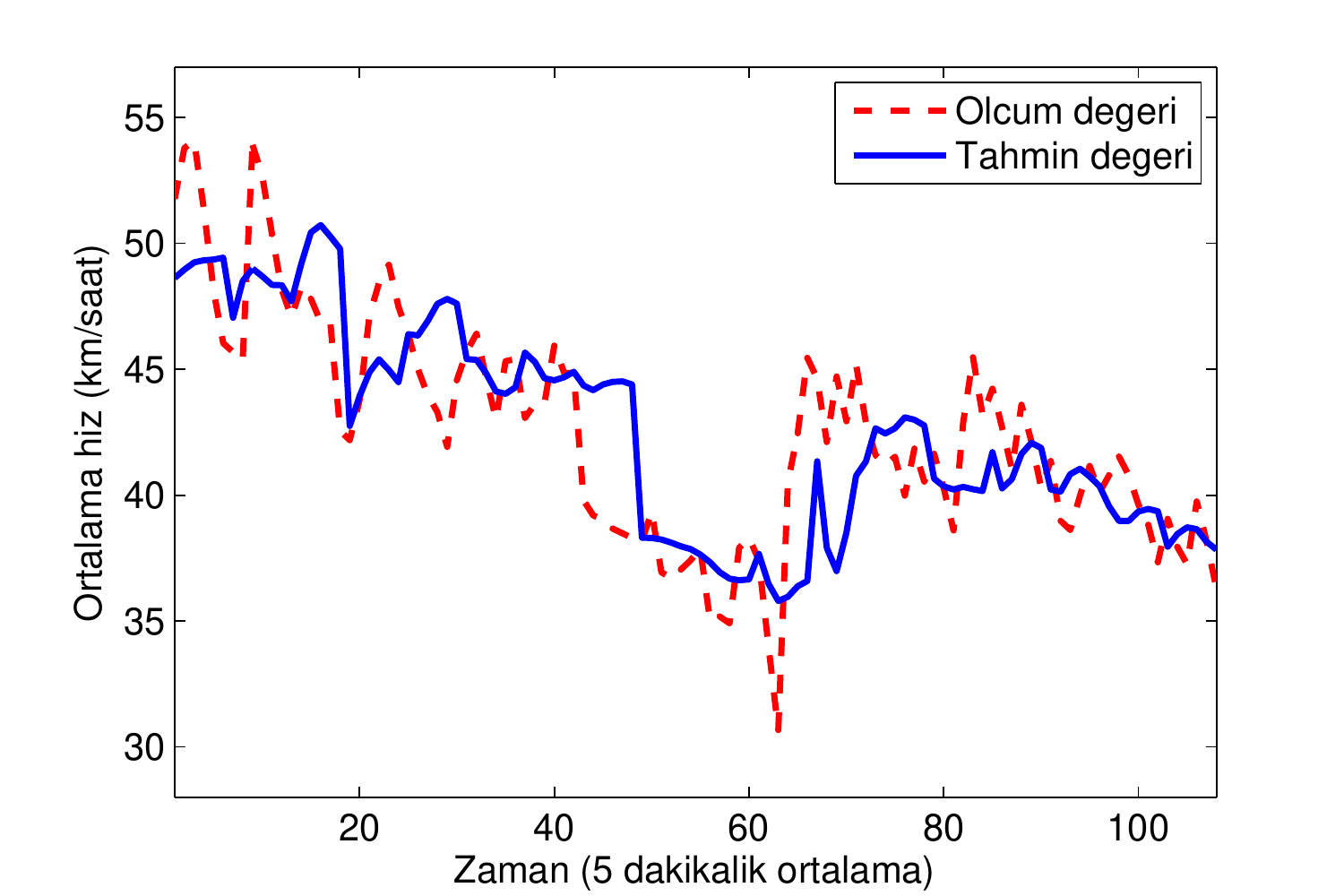}}}
		\label{AR_model}}
	\subfigure[YSA Modeli]{
		\resizebox{.45\textwidth}{!}{\includegraphics{{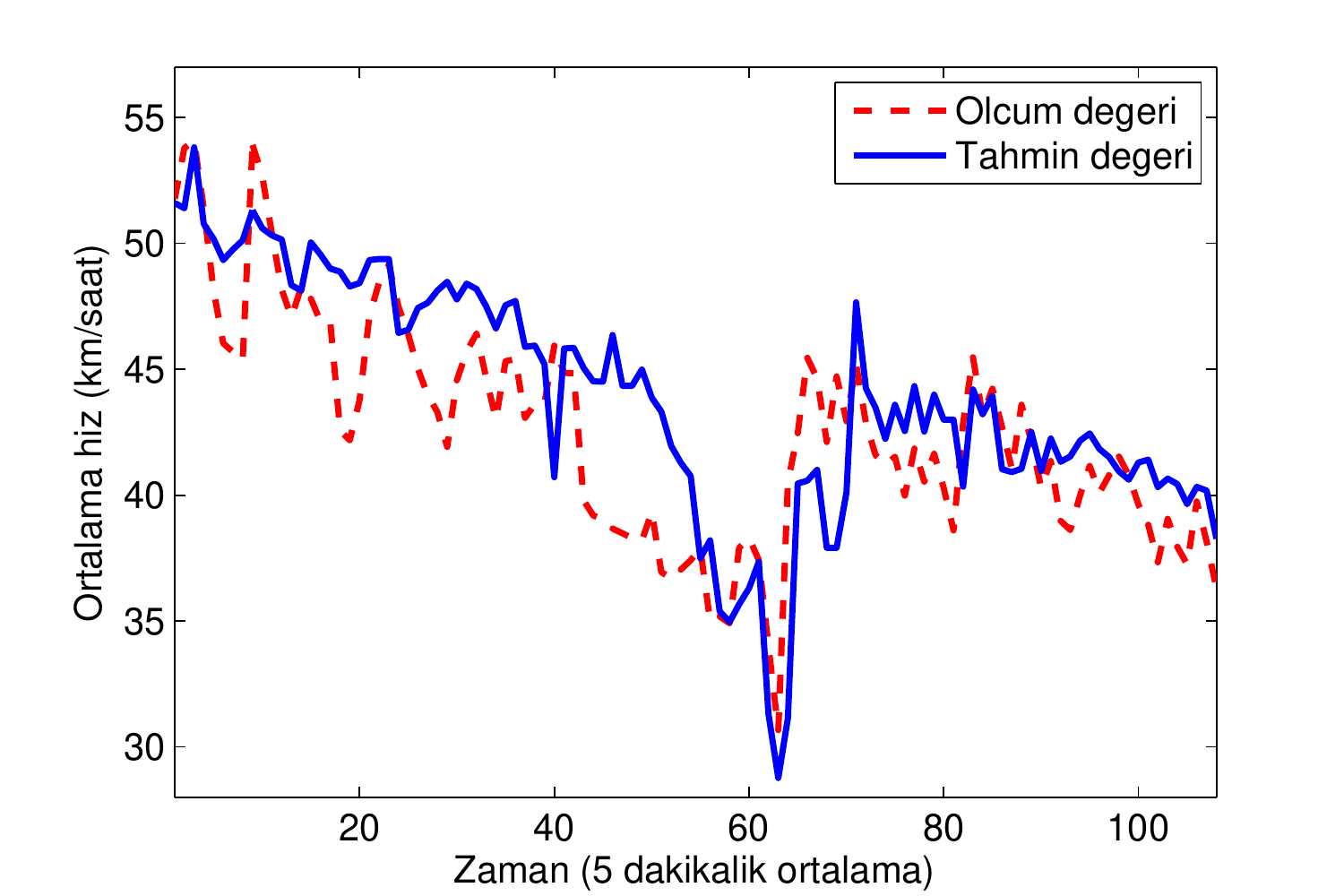}}}
		\label{YSA_model}}
	\caption{Farklı tahmin modelleri ile elde edilen tahminler.}
	\label{Karsilastirma}
\end{figure}

Önerilen yöntemin etkinliğini test edebilmek amacıyla öncelikle tahminler farklı yöntemlerle gerçekleştirilmiştir. 
Bu amaçla, önerilen modelin temelini oluşturan AR modeli ve yapay sinir ağlarına (YSA) dayanan başka bir modele ait tahminler sırasıyla Şekil~\ref{AR_model} ve \ref{YSA_model}'de verilmiştir. 
Veriler arasında doğrusal bir ilişkinin tam olarak kurulamadığı trafik akımı ve ortalama hız gibi değişkenler için literatürde genellikle YSA tabanlı yöntemler daha iyi sonuçlar vermektedirler.
Ancak bu çalışmada tüm modeller için sınırlı sayıda (108 adet 5 dakikalık veri) bir eğitim seti kullanıldığından, başarısı büyük oranda eğitim setinin boyutuna bağlı olan YSA modeli için AR modeline kıyasla gerçek değerlerden daha uzak sonuçlar elde edilmiştir.   

\begin{figure}[h!]
	\centering
	\resizebox{.45\textwidth}{!}{\includegraphics{{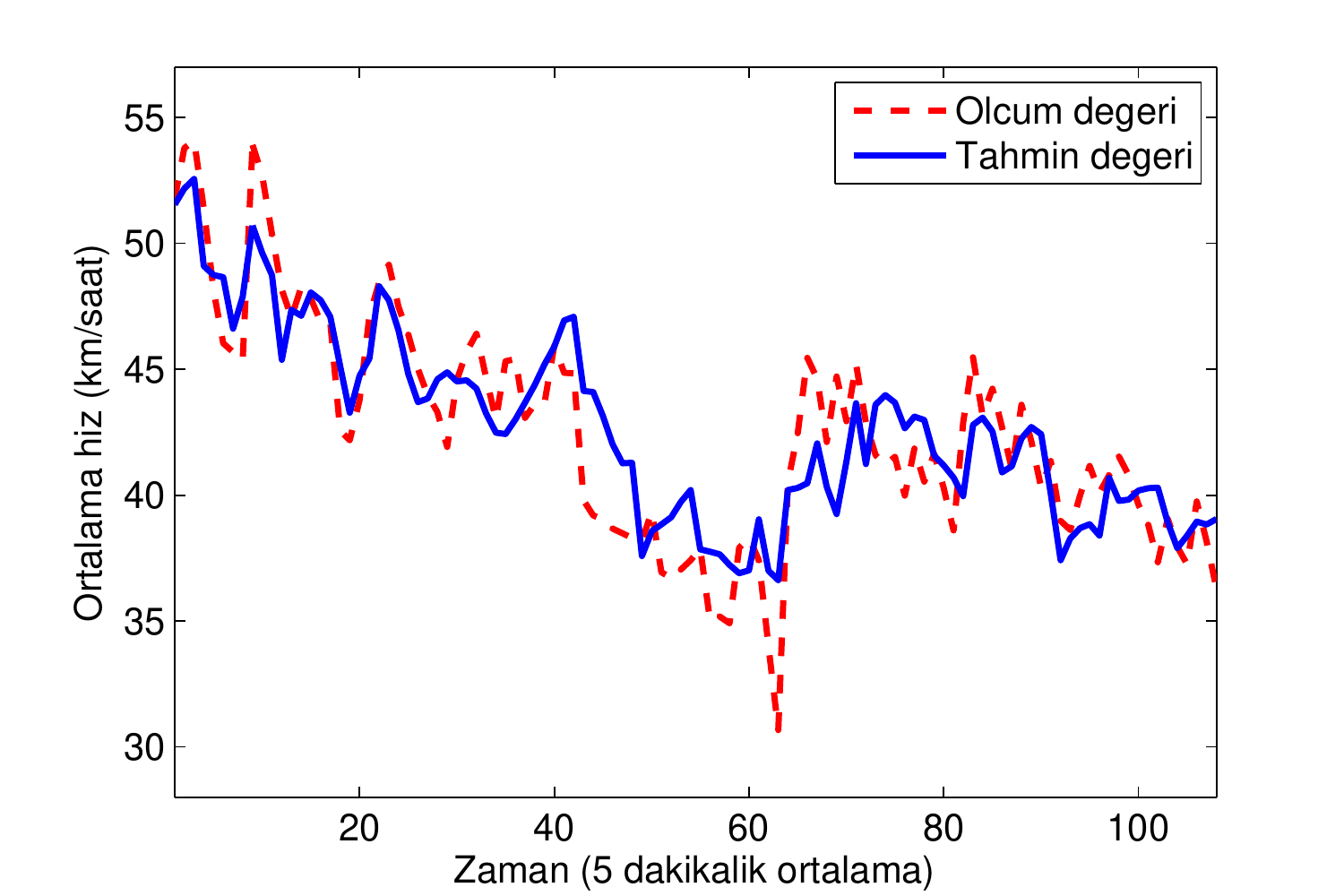}}}
	\caption{Önerilen yöntem ile elde edilen tahminler.}
	\label{CSTF}
\end{figure}

Önerilen tahmin yöntemi ile elde edilen sonuçlar ise Şekil~\ref{CSTF}'de verilmiştir.
Şekilden görülebileceği gibi yakın çevredeki noktalardan alınan ölçümlerin tahmin modeli içerisine dahil edilmesi özellikle ani değişimlerin tahmininde büyük bir fayda sağlamaktadır.
Önerilen yaklaşımın bir diğer özelliği ise zamanla değişen veri ile birlikte en yüksek modelleme doğruluğu için $x$ vektör katsayılarının sürekli olarak tekrar hesaplanmasıdır.
Başka bir ifadeyle, tahminlerin gerçekleştirildiği zamana ait olan ölçümlerin giriş verisi olarak kullanılmasıyla, her bir tahmin ufku için yeni bir $x$ vektörü elde edilmektedir ve bir sonraki tahmin ufku için gerçekleştirilecek tahminlerde bu vektör kullanılmaktadır. 

Yol üzerindeki algılayıcılardan son olarak alınan veriler ile tahminlerin sürekli olarak güncellenmesi ve bu sayede ağın akım yönüne göre başındaki ve sonundaki bağlar arasındaki trafik şartlarının değişiminin incelenmesi; sabah ve akşam zirve saatler, gece yarısı saatleri ve büyük etkinlikler (spor müsabakalar, gösteri, vb.) gibi geçici zamanlar ile kazalar ve olumsuz hava şartları gibi normal olmayan durumlar için literatürdeki benzer yöntemlere göre daha iyi sonuçlar vermektedir.
Ayrıca önerilen modelde yinelemeli (recursive) bir yaklaşım benimsenmiştir. 
Burada model eğitimi aşamasından sonraki ilk değer olan $n+M+1$ değeri tahmin edildikten sonra bu değer $n+M+2$ değerinin tahmininde kullanılmaktadır. 
Bu işlem tahmin ufku boyunca devam ettikten sonra tüm $A$ matrisi yeni ölçüm değerleri ile güncellenmektedir.

Önerilen tahmin yönteminin ve kullanılan diğer iki adet yöntemin etkinliklerinin daha iyi anlaşılabilmesi açısından yöntemler Tablo \ref{error_measures}'de çeşitli hata ölçütleri kullanılarak karşılaştırılmıştır.
Bu karşılaştırmalarda literatürde en yaygın olarak kullanılan ortalama mutlak hata (MAE), ortalama karesel hata (RMSE) ve normalleştirilmiş ortalama karesel hata (NRMSE) ölçütleri kullanılmıştır.
Özetle, MAE'nin tahminlerin istenen süre boyunca doğruluğu hakkında bilgi verdiği, RMSE'nin yüksek hataların karesini alması nedeniyle özellikle bu hataları dikkate aldığı ve NRMSE'nin tahmin edilen değişkenin genliğinden bağımsız olarak yüzde bir değer sağladığı söylenebilir. 
Tablo incelendiğinde önerilen yöntemin her üç kriter için de diğer yöntemlerden daha başarılı olduğu ifade edilebilir.

\begin{table}[h!]
	\centering
	\caption{Farklı yöntemlere ait hata ölçütlerinin karşılaştırılması}
	\small
	\begin{tabular}{lcccccc}
		\toprule
		\multirow{2}{*}{\textbf{Tahmin yöntemi}} & \textbf{MAE} & \textbf{RMSE} & \textbf{NRMSE}\\
		& \textbf{ [$km/saat$] } & \textbf{ [$km/saat$]} & \textbf{ [\%]}\\
		\midrule
		\textbf{YSA} & 2,62 & 3,25 & 13,91 \\
		\textbf{AR} & 2,31 & 2,99 & 12,79\\
		\textbf{Önerilen yöntem} &1,74 & 2,12 & 9,05\\
		\bottomrule
	\end{tabular}
	\label{error_measures}
\end{table}

\begin{figure}[h!]
	\centering
	\resizebox{.45\textwidth}{!}{\includegraphics{{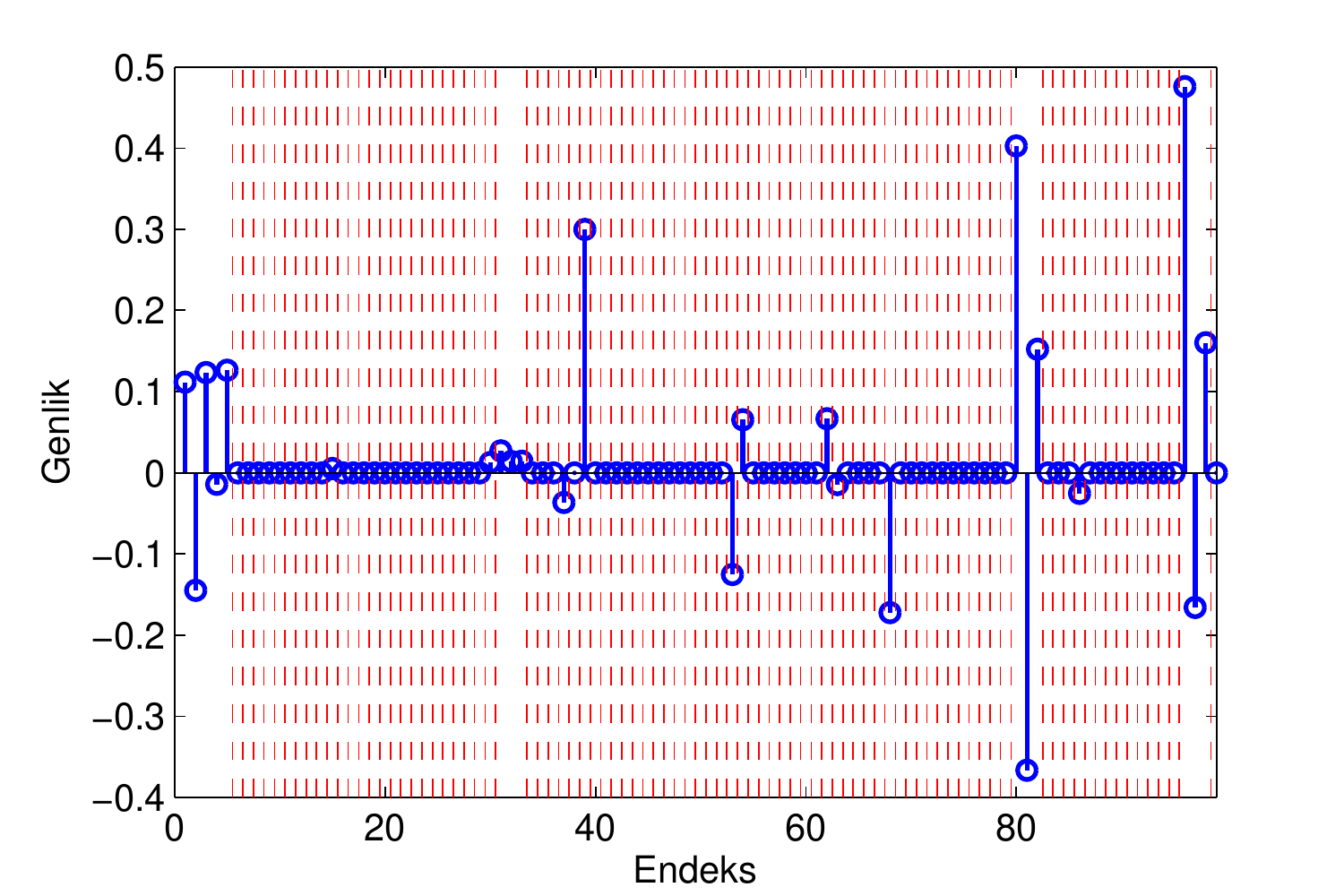}}}
	\caption{Seyrek katsayı vektörü.}
	\label{x_vector}
\end{figure}

Şekil \ref{x_vector}'de bir tahmin periyodunda (6 adet 5 dakikalik değer) önerilen modelin hesapladığı ve tahminlerde kullandığı $x$ vektörü gösterilmektedir.
Kesik çizgilerle ayrılan bölümlerden görülebileceği gibi her bir bağa ait model derecesi farklıdır.
Burada tahmin algoritmasının, daha önceden belirtildiği gibi, o andaki tahminler için yalnızca bazı noktalardaki değerleri (sıfır olmayan değerler) dikkate aldığı görülebilir.
Bu noktalar her bir yarım saatlik sürede tahmin yöntemi tarafından tekrar belirlenmektedir.
Tahmin süresi boyunca değişen $x$ vektörü incelendiğinde bazı noktaların her zaman tahmine bir katkı sağladığı, bazı noktaların ise tam aksine hiçbir zaman bir değer almadığı görülmüştür.
Verilerin alındığı internet sayfasında tüm noktaları gösteren bir harita bulunmamaktadır.
Ancak yalnızca $x$ vektörünün zamanla değişimine bakarak her bir noktanın, tahminlerin gerçekleştirildiği noktaya olan uzaklığı ve iki nokta arasındaki ilişki hakkında bir yorum yapabilmek mümkündür.
Verilerin alındığı noktaların harita üzerindeki konumlarını bilmeksizin, yalnızca veriler arasındaki korelasyonlara bakarak hangi verilerin giriş setine dahil edilebileceğine karar verebilme özelliği, önerilen yöntemin en önemli özelliklerinden biri olarak gösterilebilir.
Bu özelliği sayesinde önerilen yöntemin, özellikle eksik ve hatalı verilerin bulunduğu veri setlerinde, literatürdeki benzer yöntemlere kıyasla çok daha başarılı olacağı öngörülmektedir.

\section{Sonuçlar}

Literatürde özellikle yenilenebilir enerji kaynaklarından elde edilebilecek güç miktarlarının tahmini için son birkaç yılda oldukça fazla kullanılmaya başlayan uzam-zamansal yöntemlerin etkinlikleri göz önüne alınarak bu yayında trafiğin durumunu ifade etmek için en yaygın olarak kullanılan ortalama araç hızlarının gelecek tahminleri için bir uzam-zamansal yöntem sunulmaktadır.
Ağ üzerindeki farklı noktalardan alınan ve çok sayıda parametreye ait olan ölçümleri tahmin aşamasında model içerisine dahil ederek mevcut verilerden en üst seviyede yararlanmayı saglayan bu yöntemin, farklı noktalara ait çeşitli trafik karakteristikleri arasındaki korelasyonları kullanarak tahmin doğruluğunu önemli derecede arttırdığı gösterilmiştir. 
Trafik tıkanıklığının etkilerinin ardışık yollar üzerinde belirli bir zaman gecikmesiyle görülmesi, trafik tahminlerinde bu yöntemlerin etkin olmasını sağlamıştır.
Literatürde en iyi girişlerin bir ön analiz ile belirlendiği tahmin yaklaşımlarının aksine bu yayında önerilen yöntem her bir tahmin ufku için yalnızca bir önceki tahminde en iyi sonuçları veren değişkenleri ve değerleri model içerisine dahil etmektedir.
Bu nedenle, bir bölgedeki trafik akımındaki değişen trafik şartlarının ve hava koşullarının etkilerinin tahmin aşamasında dikkate alınması sağlanmıştır.
Çeşitli tahmin yöntemleri ile yapılan karşılaştırmalar önerilen yöntemin etkinliğini göstermektedir. 
Önerilen modelin etkinliğini arttırmak amacıyla giriş veri setinin yolculuk süreleri, meşguliyet ve kuyruklanma uzunluğu gibi trafik karakteristikleri ile sıcaklık ve yağış miktarı gibi hava değişkenleri kullanarak genişletilmesi yakın gelecekteki bir çalışma olarak planlanmaktadır.

\section{Teşekkür}

Bu çalışma Türkiye Bilimsel ve Teknik Araştırmalar Kurumu (TÜBİTAK) tarafından 115E984 no'lu "Kent içi yol ağlarının modellenmesi, kalibrasyonu ve yeni bir mikrosimülatörün yaygın olarak kullanılan simülatörlerle karşılaştırılması" projesi kapsamında desteklenmektedir.


\bibliographystyle{IEEEbib}
\bibliography{reference_TOK_2016.bib}

\end{document}